\documentclass[prb,twocolumn,showpacs]{revtex4-1}
\usepackage{epsfig}
\usepackage{dcolumn}
\usepackage{graphicx}
\usepackage{subfigure}
\usepackage{epsf}
\usepackage{color}
\usepackage{soul}
\usepackage[T1]{fontenc}
\usepackage[latin9]{inputenc}
\usepackage{array}
\usepackage{textcomp}
\usepackage{multirow}
\usepackage{amsmath}
\usepackage{graphicx}
\usepackage{esint}
\makeatletter
\begin{document}

\title{Giant electroresistance and tunable  magnetoelectricity  
in a 
multiferroic
junction}

\author{Francesco Ricci, Alessio Filippetti, and Vincenzo Fiorentini}

\affiliation{CNR-IOM (UOS Cagliari) and Department of Physics, Cagliari University, 09100-Cagliari, Italy}
\begin{abstract}
First-principles density functional calculations  show that 
 the $\textrm{SrRuO}_{3}/\textrm{PbTiO}_{3}/\textrm{SrRuO}_{3}$
multiferroic junction with asymmetric (RuO$_{2}$/PbO and TiO$_{2}$/SrO)
interfaces has a large ferroelectric depolarizing field, whose switching changes the interface transmission probabilities  for tunneling electrons, leading
to electroresistance modulation over several orders of magnitude. The  switching  further affects the interface spin density,   naturally driving magnetoresistance  as well as modulated spin-dependent in-plane
resistivity, which may be  exploited in field-effect devices. 
\end{abstract}
\pacs{73.40.-c,
77.80.Fm,        
85.50.-n,         
75.70.-i           
}
\maketitle
\section{introduction and method}
Multiferroic junctions are stacks of metallic ferromagnets and insulating
ferroelectrics where electroresistance
and magnetoresistance modulation\cite{key-1,key-3} have been obtained via polarization switching in the ferroelectric interlayer. Ferroelectric (FE) materials --especially perovskite oxides-- are used  as tunnel barriers between metal electrodes. Polarization charges accumulate at the interfaces of the finite FE layer with the rest of the stack, and cause a depolarizing field in the FE. Despite the strong screening by the metal electrodes, a sizable  field survives  in typical junctions. This persistence is the basis for nanoscale device concepts for data storage.\cite{key-23,key-38,key-39} A large (``writing'') external bias across the FE is used to switch the FE depolarizing field and polarization, and a small (``reading'') bias is then used to read the resistance of the stack in the newly realized state. This resistance may --in fact, in the present case, does-- depend  on polarization (i.e. depolarizing field) orientation, for instance because of interface structure or of  asymmetries in potential profile along the junction and the associated tunneling
probability. If that is the case, an electroresistance effect is realized.

When ferromagnetic (FM) electrodes are added to the junction, a multiferroic tunnel junction (MFTJ) is realized. Electron tunneling from the electrode through the FE barrier is now spin-dependent, and the tunneling current also depends on the relative orientation of magnetization of the two electrodes, or on the local induced magnetization. Thus, in MFTJs, tunneling magneto- and electro-resistance (TMR and TER) effects coexist, leading to four distinct states accessible via electric and magnetic external fields.\cite{key-26,key-27} 

Further, because of the same asymmetries, the interface polarization charge is spin-polarized to a degree depending on polarization direction.
 MFTJs may thus exhibit   interfacial magneto-electricity (ME), i.e.  changes
in interface magnetization induced by FE polarization reversal, hence ultimately driven by an electric field.\cite{key-28,key-17} The polarization switching   changes  sign and value of the FE charge at a given interface, but also its relative majority or minority spin content, establishing a tunable interface magnetization. Of course, this will affect both the tunneling (typically ballistic) and the in-plane (typically diffusive) transport in either spin channel.

In this paper, we study with  first-principle calculations a $\textrm{SrRuO}_{3}/\textrm{PbTiO}_{3}/\textrm{SrRuO}_{3}$ (SRO/PTO/SRO) multiferroic tunnel junction, and  specifically its tunneling electroresistance, interfacial ME coupling, and in-plane transport. Due to the chemical asymmetry of the interfaces and the strong polarization of  the FE layer, the electroresistance modulation is up to two orders of magnitude larger than in previous studies on  BaTiO$_{3}$-based 
MFTJs.\cite{key-26,key-20} We also find a  smaller but potentially useful tuning of in-plane resistance, originating from polarization-induced magnetization changes. Interface ME  is present, with coupling coefficients similar to other MFTJs.
\cite{key-21} 

The electronic and atomic structure of the SRO/PTO/SRO junction is
calculated within  density functional theory in the generalized gradient (GGA) approximation and the projector augmented wave (PAW)
as  implemented in the VASP code.\cite{key-25} PTO has been relaxed in tetragonal symmetry, obtaining $a$=3.924 \AA, $c$=4.176 \AA,  Ti-O on-axis bonds 1.79 \AA\, and 2.38 \AA, 
FE energy gain over paraelectric  108 meV, polarization P=0.86 C/m$^{2}$.
PTO cells with this structure  and cubic SRO have been stacked keeping the in-plane lattice
constant of PTO. The two interfaces between insulator and metal, $\textrm{RuO}_{2}$/PbO
at one side and $\textrm{TiO}_{2}$/SrO at the other, are labeled
 `Ru' and `Ti' below and  are simulated in an in-plane 2$\times$2 section.  
The supercell has 7 layers of PTO  and 6 layers of SRO as short-circuited ferromagnetic electrode, for a total of  260 atoms.
 We consider the  two ferroelectric states of PTO with  polarization
P pointing in opposite directions, perpendicular to the interfaces; all the quantities pertaining to  P pointing from RuO$_{2}$/PbO
to the TiO$_{2}$/SrO (``Ru to Ti'') are depicted in blue, while those for  P pointing from  TiO$_{2}$/SrO to RuO$_{2}$/PbO (``Ti to Ru'') are in red. All
configurations are reoptimized in length and relaxed with force tolerance 40 meV/\AA, using a 4\texttimes{}4\texttimes{}1 k-point Monkhorst-Pack mesh.

\section{Results}
\subsection{Charge and potential}
\label{chapot}
To analyze  the total charge density (built adding  narrow Gaussian charges at the  ions location to the electronic charge)  and electrostatic potential in the junction in the two polarization states, we feed their average over the sectional area $A$=4$a^2$ of our 2$\times$2 planar cell to a
one-dimensional square-wave  filter to obtain the  macroscopic average\cite{key-6}
\begin{equation}
\bar{\bar{n}}(z)=\frac{1}{aA}\intop_{z-a/2}^{z+a/2}dz'\intop_{A}n(x,y,z')\,dx\,dy.
\end{equation}
Similarly to Ref.\onlinecite{key-2}, we extract the monopole component of the macroscopically averaged  density as 
\begin{equation}
\frac{1}{2}
[\bar{\bar{n}}^{\rightarrow}(z-z_{0})\!-\!\bar{\bar{n}}^{\leftarrow}(z_{0}-z)]\, 
\end{equation}
combining  the density
profiles $\bar{\bar{n}}^{\rightarrow}$ and $\bar{\bar{n}}^{\leftarrow}$
 for  the two P states ($z_{0}$ is chosen to minimize the monopoles and ends up near the midpoint of the FE layer). 

\begin{figure}[h]
\begin{centering}
\includegraphics[scale=0.22]{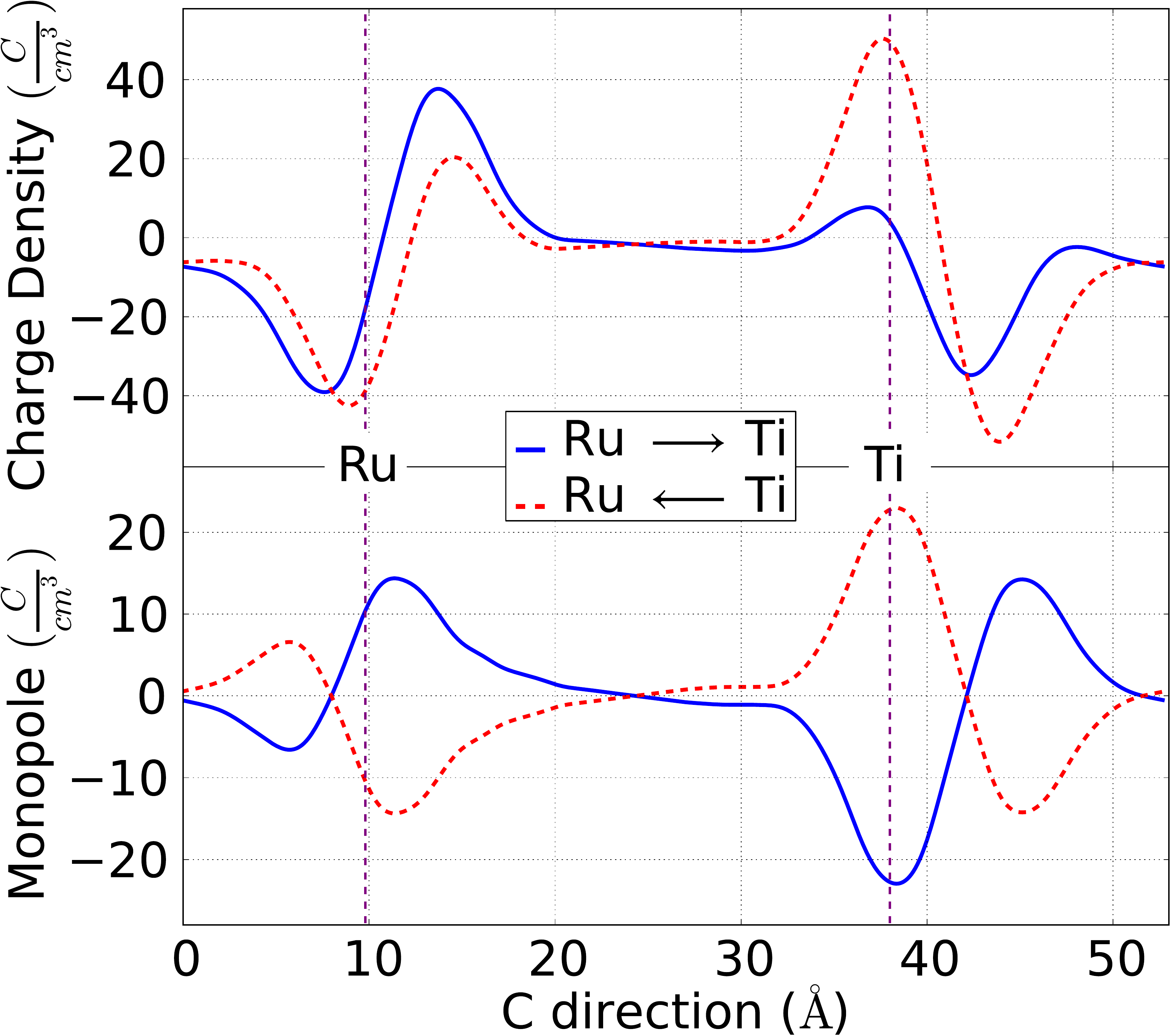}
\par\end{centering}
\centering{}\caption{\label{fig:total-charge-monopole}(Color on line) Macroscopic average  (top) of the total  charge density,  and its  monopole component (bottom) 
for the two directions of P in PTO.}
\end{figure}

\begin{figure}[ht]
\begin{centering}
\includegraphics[scale=0.24]{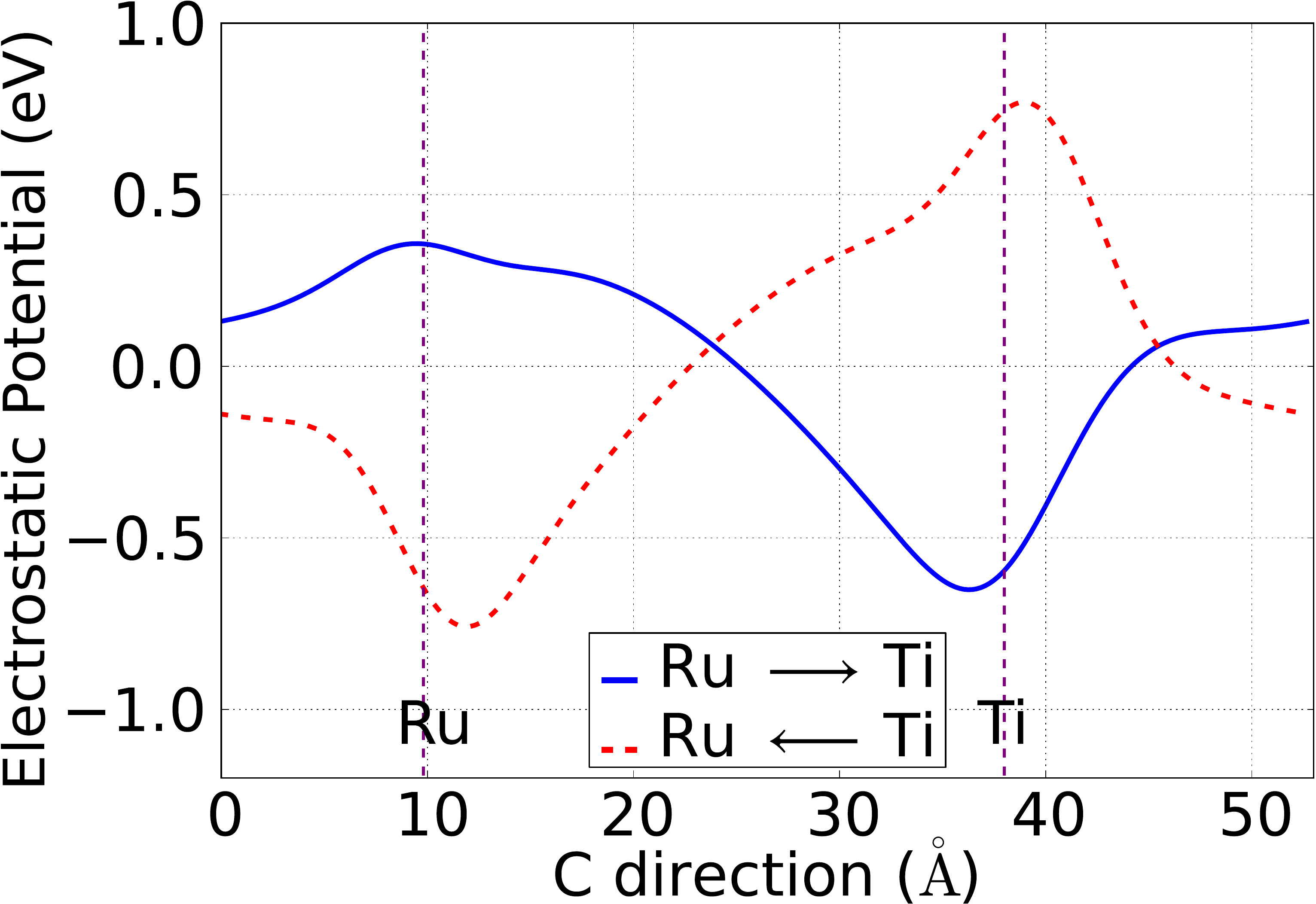}
\par\end{centering}
\caption{\label{fig:Electrostatic-potential-profile}(Color on line) Asymmetric electrostatic potential
profiles along $c$ for opposite P's in 
PTO. The depolarizing  field is $\sim$5$\times$10$^{8}$ V/m.}
\end{figure}

\begin{figure}[h]
\begin{centering}
\includegraphics[scale=0.9]{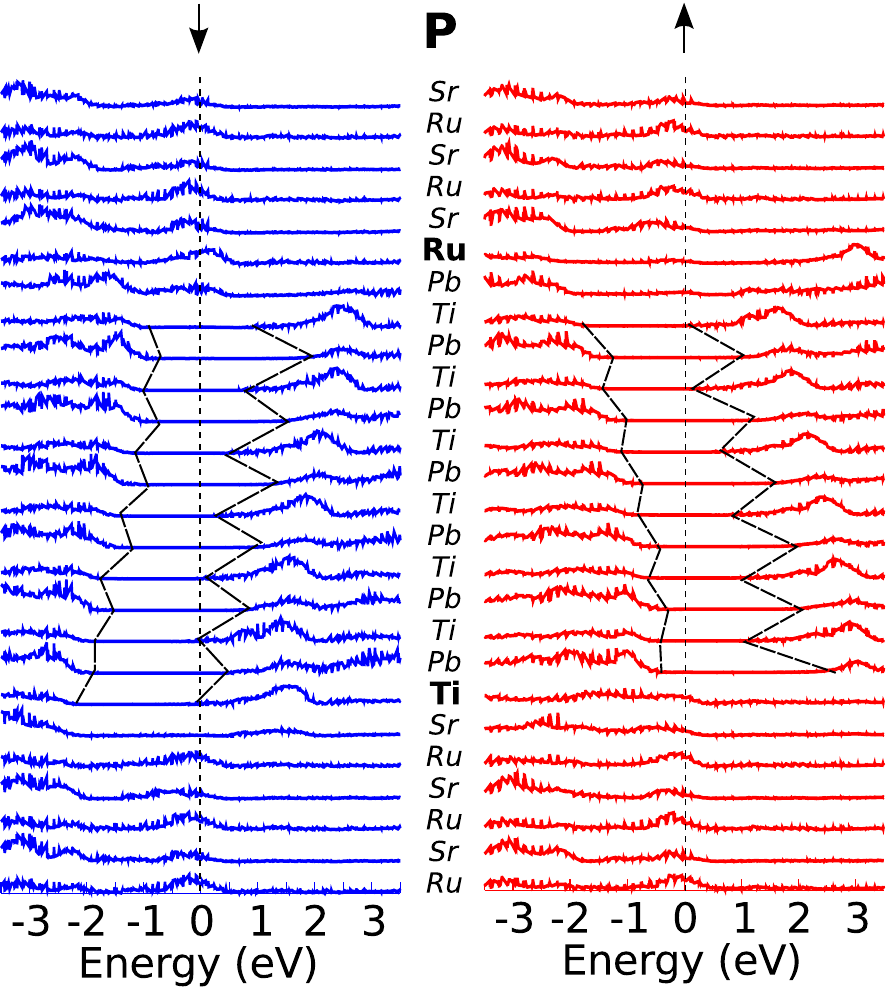}
\par\end{centering}
\centering{}\caption{\label{fig:dos}(Color on line) LDOS for the two polarization states.}
\end{figure}

The results are shown 
in Fig.\ref{fig:total-charge-monopole}.
Firstly, there is a significant net accumulation of  charge at the
two interfaces, due to the interface polarization discontinuity charge screened by the metal electrons and by the ionic response in both the insulator and the metal; the monopole is quite delocalized in comparison to that at semiconductor interfaces.\cite{key-2} This local charge produces  a depolarizing field $E_{\rm dep}$$\sim$5$\times$10$^{8}$ V/m in the FE layer, which can be read off  the potential profile in Fig.\ref{fig:Electrostatic-potential-profile}, as well as from the local density of states (Fig.\ref{fig:dos}) discussed below. For a finite polarized PTO slab of this thickness, the  expected depolarizing field would be about 9$\times$10$^{8}$ V/m assuming\cite{dc} a static dielectric constant $\varepsilon_s$$\sim$100. The large  residual  field indicates that  the metallic screening is incomplete, as suggested
earlier.\cite{key-23,key-26}

Secondly, the asymmetric interfaces of highly polarized PTO produce  quite different potential profiles
for the two  polarization directions, without any symmetry-breaking layer of other materials  interposed between electrode and barrier.\cite{key-20} This implies that the tunneling resistance along the
junction will be changed by the switching of PTO polarization. As
suggested  by simplified models,\cite{key-14} for such asymmetric potential and large residual electric field one expects a strong TER effect, which we now demonstrate calculating a) the semiclassical tunneling conductance through
the 1D potential profile of the junction, and b) the transmission coefficient
from the evanescent-wavefunction ratio in the insulator.

\subsection{Tunneling electroresistance: WKB}

\begin{figure}[ht]
\begin{centering}
\includegraphics[scale=0.93]{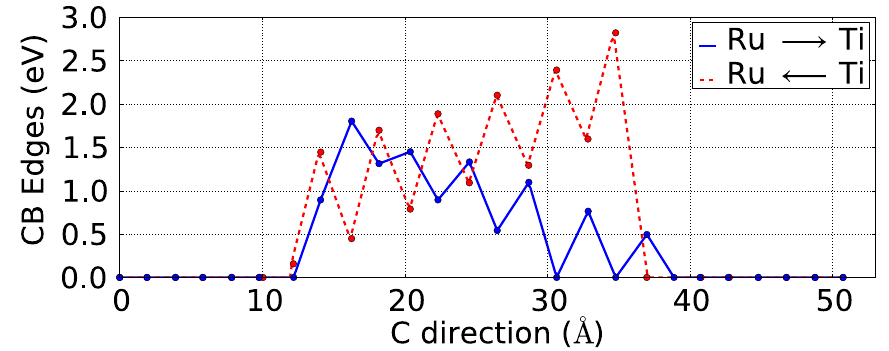}
\par\end{centering}
\centering{}\caption{\label{fig:edgeprof}(Color on line) CB edge potential profile for tunneling.}
\end{figure}

To  quantify  the  TER in our junction,
we first study the semiclassical tunneling conductance through
the 1D potential profile of the junction. An appropriate model for the potential profile through which the electrons tunnel is  the position-dependent conduction band edge of the junction referred to Fermi energy $E_F$. To extract this profile, we calculate for the two polarization orientations the layer-resolved local density of states (LDOS), which is displayed in Fig.\ref{fig:dos}. The position dependent PTO band edges shift along the junction at a rate determined by the depolarizing field, whose value is in the mid 10$^8$ V/m as already indicated by the averaged potential profile. We then extract from the LDOS the position of the conduction edge in each layer, and use it to construct the potential profiles for the two values of P, which, as shown in Fig.\ref{fig:edgeprof}, are strongly asymmetric.

\begin{figure}[h]
\begin{centering}
\includegraphics[scale=0.93]{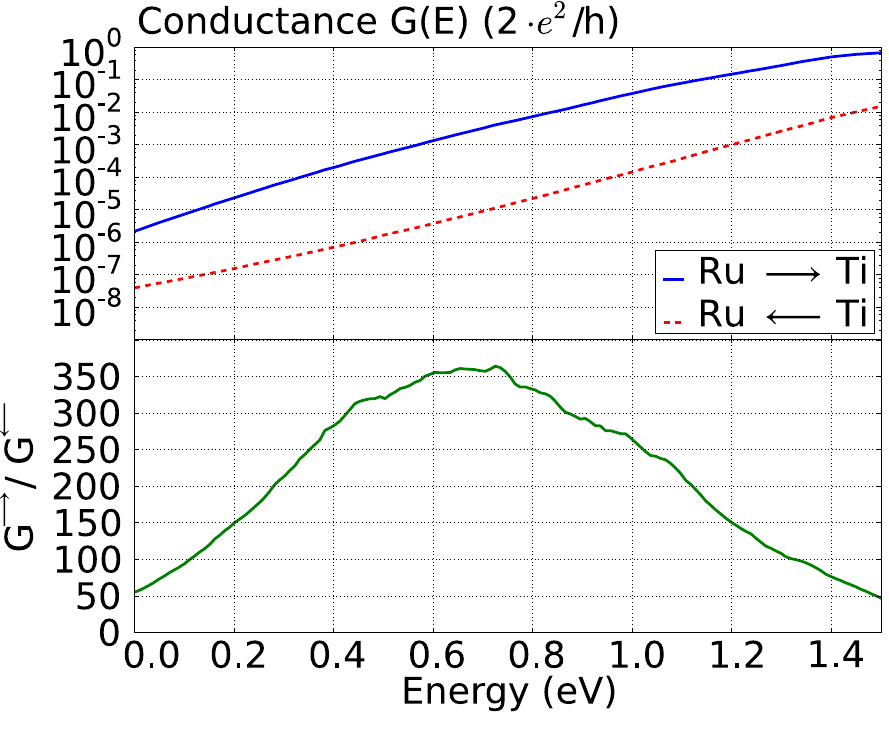}
\par\end{centering}

\centering{}\caption{\label{fig:ger}(Color on line) Top: conductance in the WKB approximation for tunneling through the 1D potential profiles
shown in Fig. \ref{fig:edgeprof}. Bottom: the TER between the two P states, topping at about  350.}
\end{figure}

We then calculate the tunneling probability 
 in the semiclassical Wentzel-Kramers-Brillouin  approximation through these two  profiles vs. injection energy, choosing the Fermi energy as zero. (This describes  electrons tunneling into the PTO conduction band. Hole tunneling is neglected due to   the large barriers and effective masses.)  As shown in Fig.\ref{fig:ger} the ratio $G_{\rightarrow}$/$G_{\leftarrow}$, i.e. the TER, is between 50 and 350 depending on energy, and therefore up to two orders of magnitude larger than in BaTiO$_3$/SRO junctions.\cite{key-26} The absolute values of $G$ are comparable
with those found for  similar junctions. We note that if we roughly estimate the writing voltage needed to reverse the depolarizing field in this structure as the field times the PTO thickness we find $E$$\cdot$$d$$\sim$2 V; thus, it would be safe to use a standard\cite{chou} reading voltage of 0.5-0.6 V, which would yield  a near-maximum TER.

\subsection{Tunneling electroresistance: Transmission function}
The second indication of a giant TER effect comes from  the two transmission functions $T^{\rightarrow,\leftarrow}\left(\mathbf{k}_{||}\right)$
across the two interfaces in the two poling directions. For not too
thin a barrier, the transmission function of a tunnel junction could
be factorized\cite{key-15} as 
\begin{eqnarray}
\begin{array}{c}
T^{\rightarrow}\left(\mathbf{k}_{||}\right)=t_{\rm Ru}^{\rightarrow}\left(\mathbf{k}_{||}\right)\exp\left(-2\kappa\left(\mathbf{k}_{||}\right)d\right)t_{\rm Ti}^{\rightarrow}\left(\mathbf{k}_{||}\right)\\
T^{\leftarrow}\left(\mathbf{k}_{||}\right)=t_{\rm Ru}^{\leftarrow}\left(\mathbf{k}_{||}\right)\exp\left(-2\kappa\left(\mathbf{k}_{||}\right)d\right)t_{\rm Ti}^{\leftarrow}\left(\mathbf{k}_{||}\right)
\end{array}
\end{eqnarray}
where $d$ is the barrier thickness, $\kappa\left(\mathbf{k}_{||}\right)$
is the lowest decay rate in the barrier and   $t_{\rm Ru,Ti}^{\rightarrow,\leftarrow}$ are the transmission
probabilities from the left or right electrode into the barrier across
the interfaces (Ru or Ti respectively) for an electron with a given
$\mathbf{k}_{||}$, for both polarization directions.

\begin{figure}[h]
\begin{centering}
\includegraphics[scale=0.22]{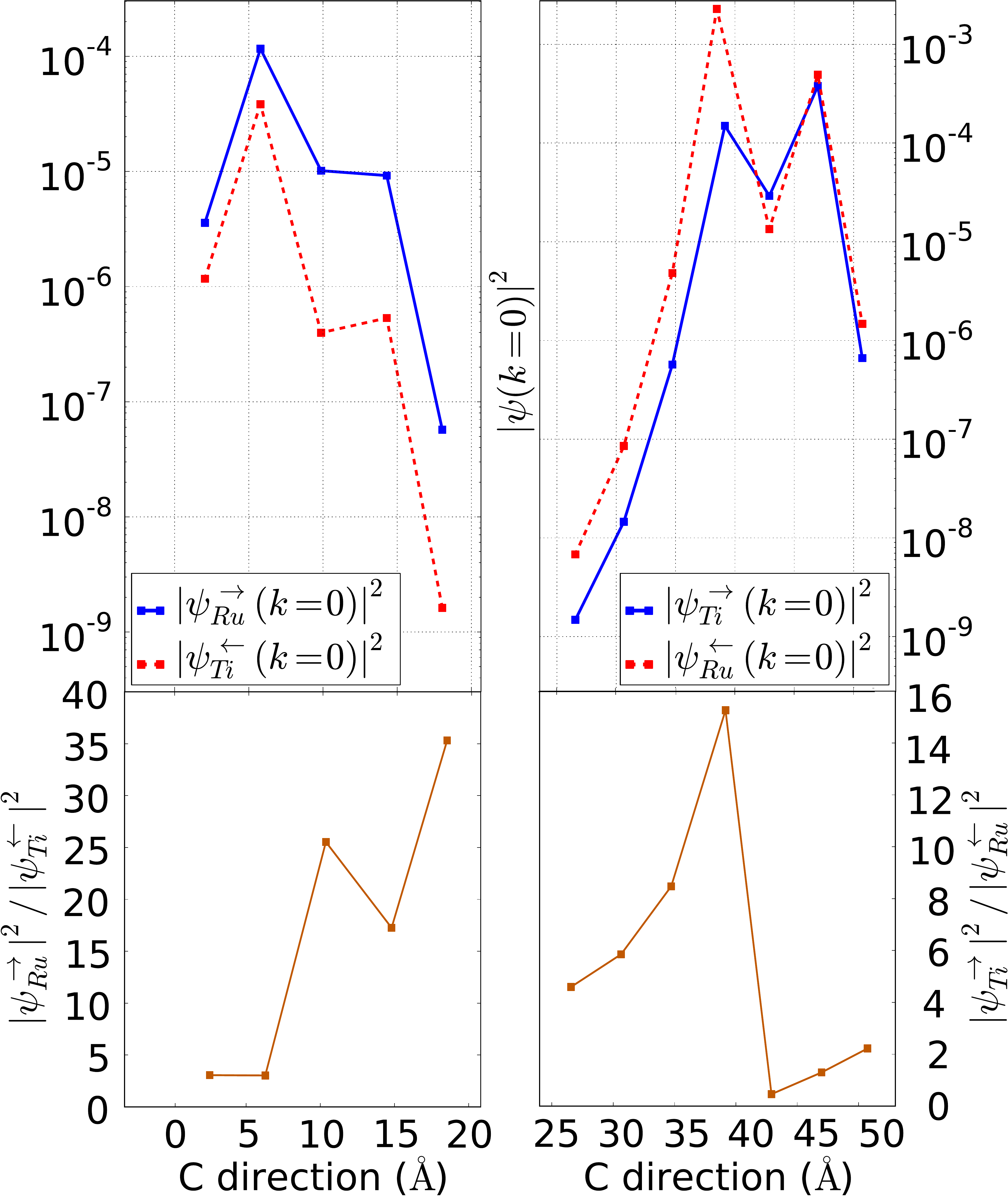}
\par\end{centering}
\centering{}\caption{(Color on line) Probability densities at the   Fermi energy and {\bf  k}$_{||}$=0 (top) and their ratio (bottom)  for the two polarization directions near the interfaces. Values taken at Ti sites}
\label{fig:wfc-ratio}
\end{figure}

 In our junction,
the two interfaces  differ both chemically and electrostatically depending on the direction of polarization, hence  we have four different $t$'s to assess. Assuming the same exponential decay 
for the two polarizations, 
we have 
\begin{equation}
\frac{T^{\rightarrow}}{T^{\leftarrow}}=\left(\frac{t_{\rm Ru}^{\rightarrow}}{t_{\rm Ti}^{\leftarrow}}\right)\cdot\left(\frac{t_{\rm Ti}^{\rightarrow}}{t_{\rm Ru}^{\leftarrow}}\right).
\end{equation}

The two $t$ ratios can be extracted from the ratio of the density
of metal-induced gap states at distance $z$ from the two interfaces
for the two poling directions, i.e.
\begin{equation}
\left|\psi_{\rm Ru,Ti}^{\rightarrow,\leftarrow}\right|^{2}\propto t_{\rm Ru,Ti}^{\rightarrow,\leftarrow}\left(\mathbf{k}_{||}\right) \exp\left[-2\kappa\left(\mathbf{k}_{||}\right)z\right].
\end{equation}
The upper panels of   Fig.\ref{fig:wfc-ratio} shows the wavefunctions as function of $z$
for $\mathbf{k}_{||}$=0 at the Fermi energy along the junction for the
two polarization directions, and the lower panels  display their ratios. Thus, $t_{\rm Ru}^{\rightarrow}$/$t_{\rm Ti}^{\leftarrow}$$\sim$25
and $t_{\rm Ti}^{\rightarrow}$/$t_{\rm Ru}^{\leftarrow}$$\sim$10, so that   
the transmission ratio $T^{\rightarrow}/T^{\leftarrow}$ is about 250, a measure of the TER comfortingly similar   to the WKB result shown above.

\subsection{Interface barriers and the origin of asymmetry}

The  asymmetry giving rise to the large TER is of electronic origin. This can be seen from the LDOS in Fig.\ref{fig:dos}: at the  Ru-Sr-Ti interface  (bottom right in Fig.\ref{fig:dos}), the PbO layer opposes a  large barrier to tunneling, while the TiO$_2$ layer adjacent to SRO is metallized;  at the Ru-Pb-Ti interface (top left in Fig.\ref{fig:dos}) the PbO layer  in contact with SRO  is metallized, and the first barrier is the smaller one provided by the TiO$_2$ layer. Put differently,  the conduction Schottky barriers between SRO and  PTO for the two interfaces are  different; indeed, our estimated difference in the conduction edge position  at the interface agrees  with the calculated\cite{umeno} 0.7 eV difference between Schottky barriers of the two interfaces.\cite{nota} Since despite the gap underestimate in GGA this difference is well reproduced, both the TER and the absolute tunneling conductance should be considered quite accurate. We conclude that the smaller TER asymmetry in SRO/BaTiO$_3$ is related to its lesser or absent Schottky barrier asymmetry.\cite{jg} 

\begin{figure}[h]
\begin{centering}
\includegraphics[scale=0.96]{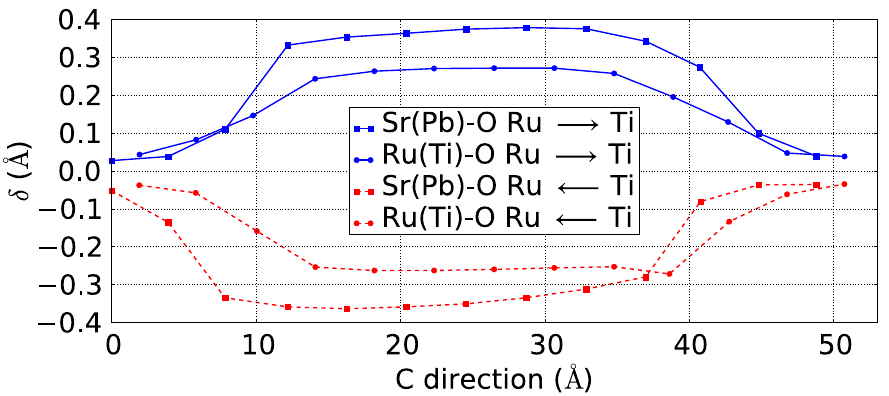}
\par\end{centering}
\centering{}\caption{\label{fig:disp}(Color on line) 
Cation displacement with respect to the plane of surrounding oxygens.}
\end{figure}

Ionic screening, in turn, presents no significant surprise:  cationic displacements in the junction  are rather well behaved,  as shown by Fig.\ref{fig:disp}; they agree largely  with previous estimates\cite{umeno} and are fairly similar to BTO/SRO (although here they are larger on average consistently with the  larger polarization of PTO).

\subsection{Interface magnetoelectricity}

We now consider  the interface ME effect.
We elect to estimate the interface ME coefficient $\alpha$ assuming a linear magnetization-field relation  
\begin{equation}
\mu_{0}\Delta M=\alpha E.
\label{MErel}
\end{equation} 
Given the likely importance of non-linearity for the high fields involved here, this should be considered an order-of magnitude estimate.
There is some latitude in deciding which magnetization changes are to be considered, depending on the operational procedure or application  envisaged. 
In the present context, the natural scenario is  polarization switching: $\Delta$$M$ is the integrated difference of magnetization density at each interface between the two polarization states. Clearly,  two  interface-related ME coefficients will result, either one of which will be relevant in practice depending on which interface is active in the specific experiment or application. To operationally implement this scenario, one just needs to switch P  via the writing voltage. 

\begin{figure}[h]
\begin{centering}
\includegraphics[scale=0.27]{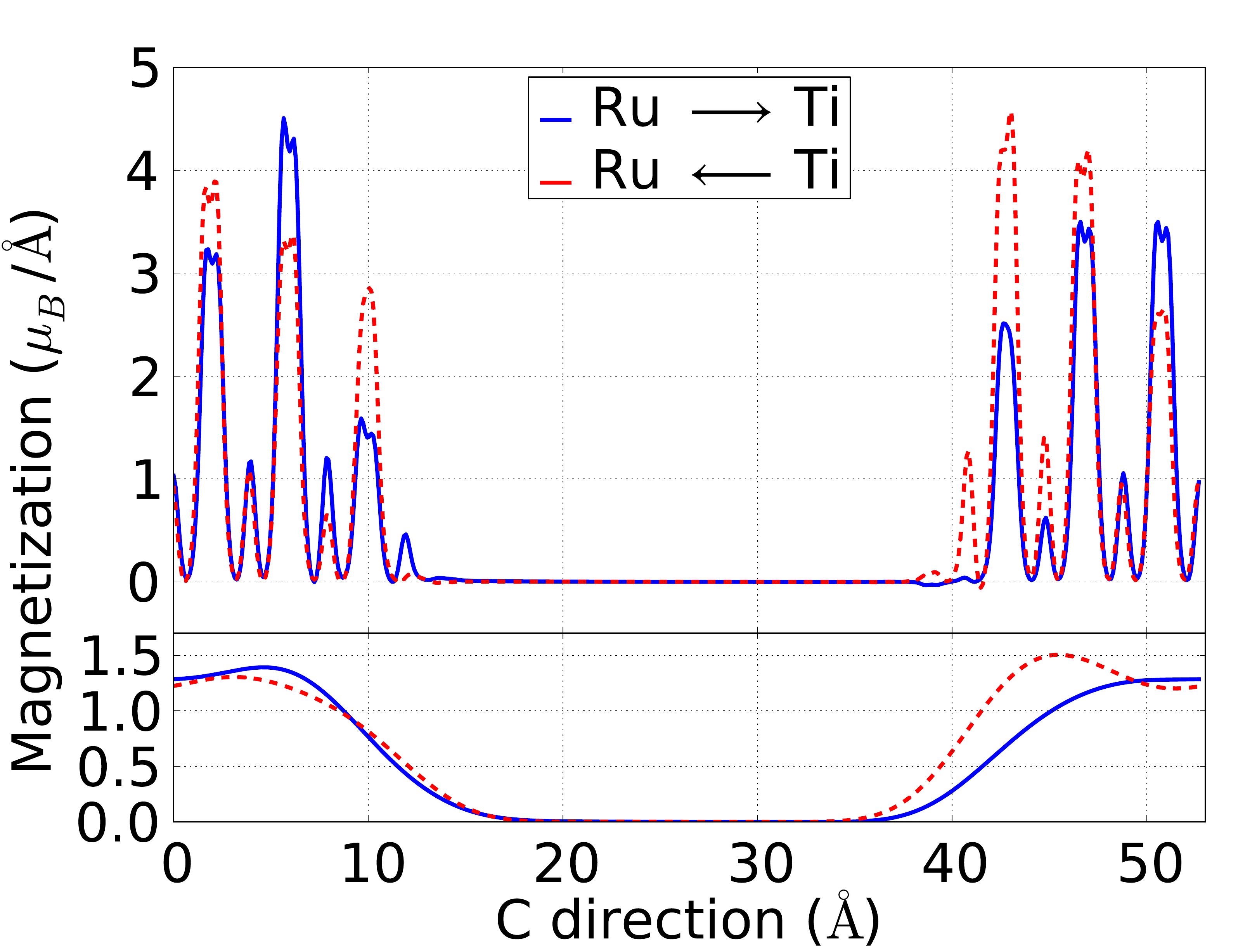}
\par\end{centering}
\centering{}\caption{\label{fig:ind-spin-density} (Color on line) Planar and macroscopic averages of the magnetization in the two P states.}
\end{figure}

To calculate the $\Delta$$M$'s  we define in analogy to the charge density in Sec.\ref{chapot}  the macroscopic averages of the 
magnetization density $\overline{\overline{m}}^{\rightarrow}$($z$) and $\overline{\overline{m}}^{\leftarrow}$($z$)  for the two states  Ru$\rightarrow$Ti and Ru$\leftarrow$Ti of PTO polarization. 
In Fig.\ref{fig:ind-spin-density} we report the planar and  macroscopic averages 
 for the two P states. Upon switching, the magnetization changes strongly at the Ti interface, much less at the Ru one.

To calculate the $\alpha$'s we conventionally  choose the initial P state to be 
Ru$\rightarrow$Ti and the final state  Ru$\leftarrow$Ti.   The magnetization changes  at the Ru or Ti interface are 
\begin{eqnarray}
\Delta M_{\rm Ru}&=&\intop_{\rm Ru} 
(\overline{\overline{m}}^{\rightarrow}-\overline{\overline{m}}^{\leftarrow})\, dz\nonumber\\
\Delta M_{\rm Ti}&=&\intop_{\rm Ti} 
(\overline{\overline{m}}^{\rightarrow}-\overline{\overline{m}}^{\leftarrow})\, dz,
\label{dM}
\end{eqnarray}
where the integrals are done near each interface between the region of zero magnetization within PTO and the region of constant magnetization within SRO, specifically (see Fig.\ref{fig:ind-spin-density}) between 2 and 21 \AA\, for the Ru interface, and 31 and 49 \AA\, for the Ti interface. We choose as   electric field in  Eq.\ref{MErel} the  depolarizing field  $E_{\rm dep}$, which is 
 taken positive by convention (a different choice  will simply change the sign of both ME coefficients). 
We then obtain the ME coefficients as
\begin{equation}
\alpha_{\rm Ru}=\frac{\mu_{0}\Delta M_{\rm Ru}}{E_{\rm dep}}, \ \ \ \alpha_{\rm Ti}=\frac{\mu_{0}\Delta M_{\rm Ti}}{E_{\rm dep}}.
\label{MEcoef}
\end{equation}
Table \ref{tab:table} summarizes the induced
magnetizations (Eq.\ref{dM}) and the ME coefficients at the two interfaces  (Eq.\ref{MEcoef}).
Again the asymmetry in the interfaces shows up dramatically. The ME coefficients are somewhat smaller than those predicted for other similar  MFTJs,\cite{key-17,key-21} despite our induced  magnetizations  being larger. This is due to our assuming conservatively  a switching  electric field larger than typical operational coercive fields of PTO. These may be 5 to 20 times smaller depending on external conditions and sample properties (and hence the $\alpha_s$ could be larger by the same factors).

\begin{table}[ht]
\caption{\label{tab:table} Magnetization changes ($\mu_{\rm B}$ per interface cell) and ME coefficients (10$^{-11}$G$\cdot$cm$^{2}$/V) upon switching P from 
Ru$\rightarrow$Ti to  Ru$\leftarrow$Ti.}
\begin{centering}
\begin{tabular}{crr}
\hline\hline
Interface & \multicolumn{1}{r}{$\Delta$$M$} &
\multicolumn{1}{c}{$\alpha$}\\
\hline 
Ru &  --0.31& --0.11\\
Ti & 4.46 & 1.62 \\
\hline\hline
\end{tabular}
\par\end{centering}
\end{table}

\begin{figure}[ht]
\begin{centering}
\includegraphics[scale=0.6]{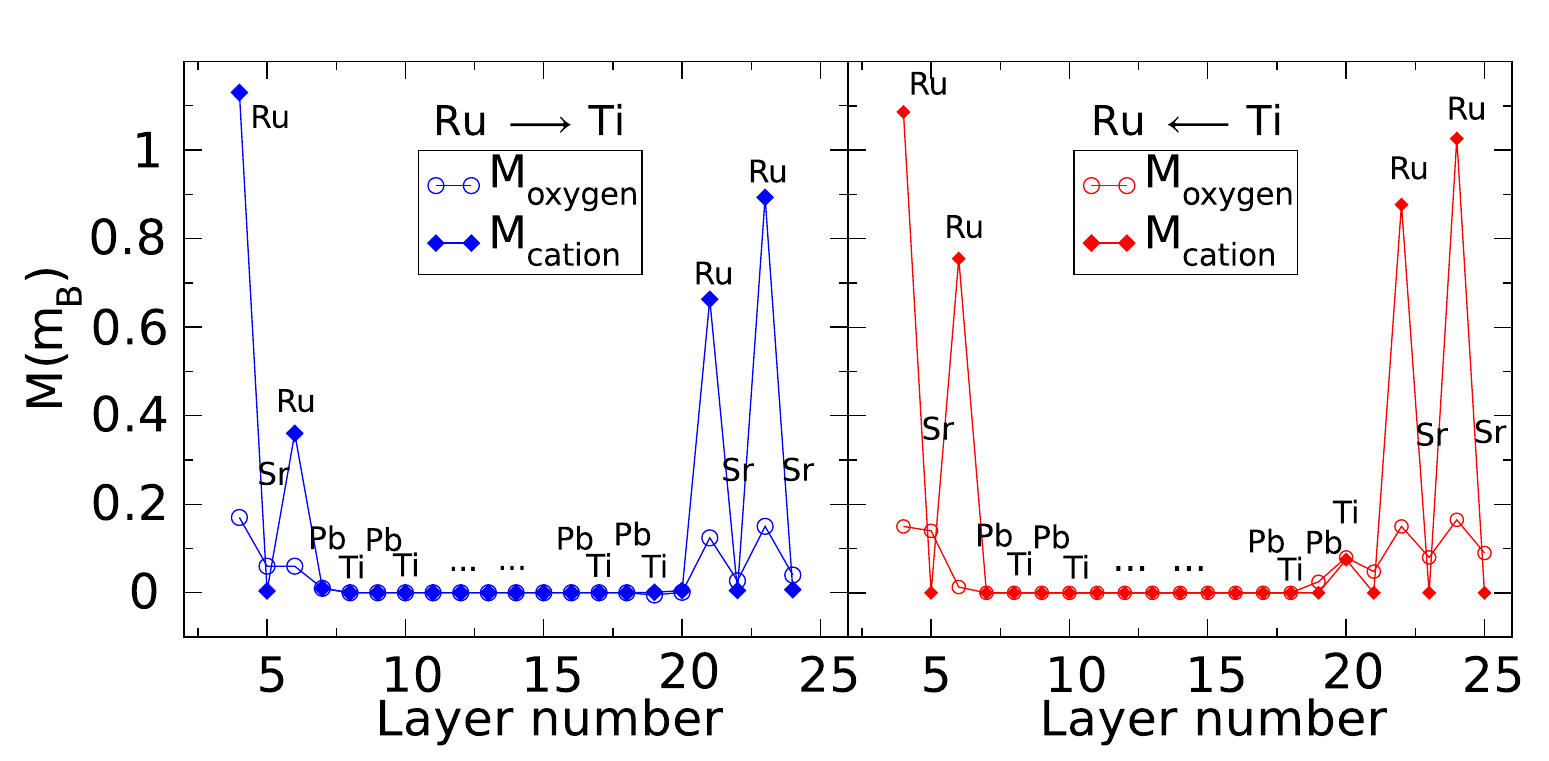}
\end{centering}
\caption{\label{fig:magint}(Color on line) Atom-resolved magnetization ($\mu_B$ per atom) for the two polarization states (a few SRO layers omitted).}
\end{figure}

Some additional insight can be gained by examining  in Fig.\ref{fig:magint} the atom-resolved average magnetic  moments  obtained integrating the magnetization density within atomic spheres defined by the PAW construction. Consistently with its larger ME coefficient,
the  Ti interface (right side of each panel) is more magnetically polarizable  than the Ru interface. For Ru$\leftarrow$Ti polarization (right panel),  magnetization builds up in the  first Ru layers,  spilling over into the Ti metallized layer and through to the first insulating PbO layer. In Ru$\rightarrow$Ti polarization (left panel), the interface Ru's  lose some (and Ti and oxygens, all) of  their moment.    At the Ru  interface (left side of each panel), the magnetization does not extend at all into PTO (specifically in the PbO layer) in either case, and the SRO magnetization mostly redistributes among Ru's and O's in the first and second layer, changing  only slightly overall (see also Fig.\ref{fig:ind-spin-density}).

In closing this Section, we note that another possible  choice for $\Delta$$M$ is the interface magnetization density change  with respect to zero electric field. In our case, this translates into the magnetization difference at the two interfaces with or without the depolarizing field, i.e. for ferroelectrically-distorted or paraelectric PTO. As  P, i.e. the field, can be turned on in two ways, this procedure  produces four ME coefficients,  
$\alpha_{\rm Ru}^{\rightarrow}$$\simeq$$\alpha_{\rm Ru}^{\leftarrow}$=0.62, $\alpha_{\rm Ti}^{\rightarrow}$=0.14, $\alpha_{\rm Ti}^{\leftarrow}$=1.75, in units of 10$^{-11}$G$\cdot$cm$^{2}$/V. 
This polarization ``turn-on'' scenario, unfortunately, is essentially impracticable. It would require  forcing PTO across its ferroelectric transition, e.g. by lowering the temperature across the Curie point  T$_c$, with a small poling voltage applied to the junction to select the desired P state. This is largely incompatible with device operation  due to the high   T$_c$$\simeq$500$^{\circ}$ C.

\begin{figure}[ht]
\begin{centering}
\includegraphics[scale=0.32]{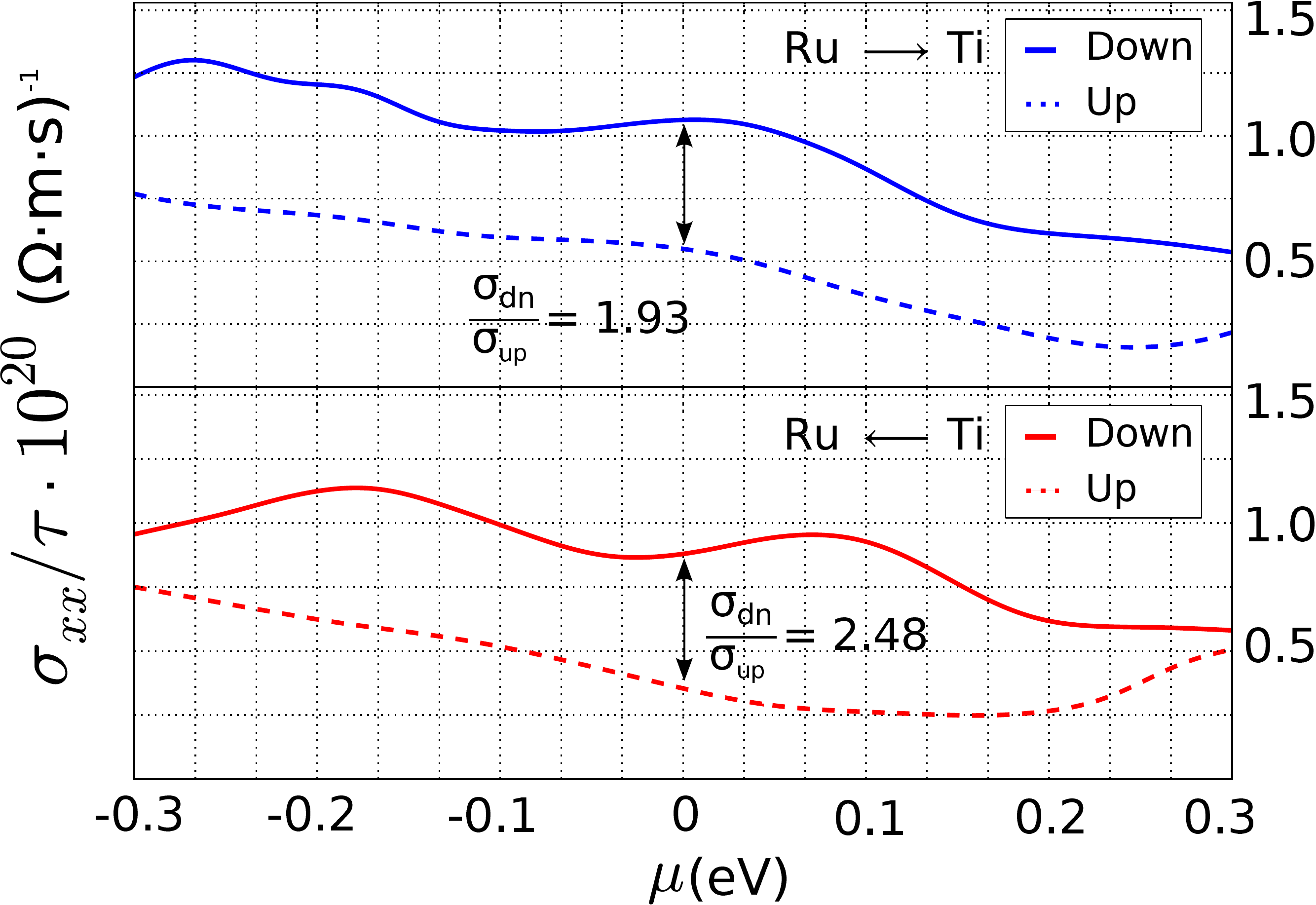}
\includegraphics[scale=0.45]{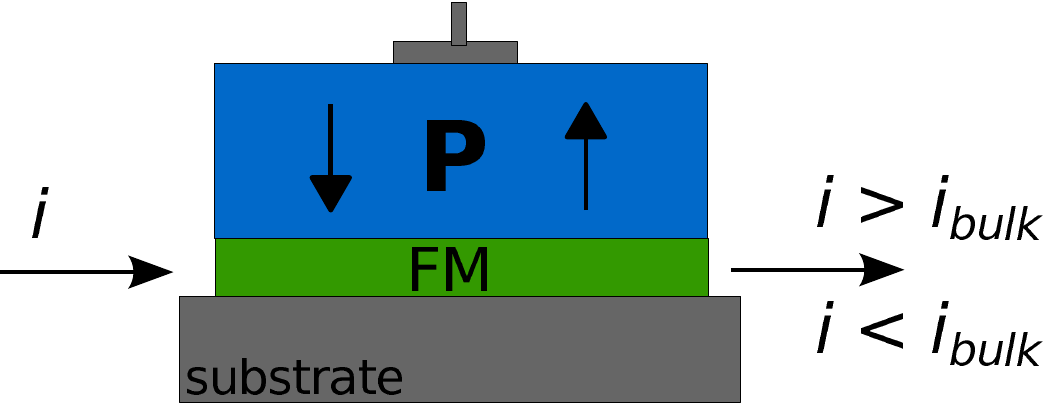}
\end{centering}
\caption{\label{fig:cond-xx-eV} (Color on line) In-plane spin-resolved conductivity vs chemical potential  for the two
 polarization states.   Bottom: schematic
device concept  exploiting conductivity modulation.}
\end{figure}

\subsection{In-plane conductivity modulation}
The  sizable interface asymmetry and  ME coupling 
will  influence the in-plane current  in the SRO layer. To assess this effect, we  use  the BoltzTraP\cite{key-16} code to calculate the conductivity in the plane of the junction in the diffusive regime as
\begin{equation}
\sigma_{\alpha\beta}\left(T;\mu\right)=\frac{1}{\Omega}\int\sigma_{\alpha\beta}\!\left(\epsilon\right)\left[-\frac{\partial f\left(T;\epsilon\right)}{\partial\epsilon}\right]d\epsilon
\end{equation}
with 
\begin{equation}
\sigma_{\alpha\beta}\left(\epsilon\right)=e^{2}\sum_{i{\bf k}}\tau_{i,\mathbf{k}}v_{\alpha}\left(i,\mathbf{k}\right)v_{\beta}\left(i,\mathbf{k}\right),
\end{equation}
 where $i$ is a band index,
$\tau$ a relaxation time, $v$ the group velocity calculated from the band structure, and $f$ the occupation function. Assuming $\tau$ constant, we can  plot $\sigma$/$\tau$ vs  chemical potential. Although the scattering mechanisms may not be described in full detail,  this approximation is quite sufficient to address conductivity ratios between polarization states. (We note in passing  that diffusive conductivity is  appropriate for in-plane transport,  but may be inapplicable to tunneling transport depending on the nature of the Fermi surface of the junction system. In the present case this approach gives a TER $\sim$10$^5$ which, while consistent  with similar calculations for 1+6 PTO/SRO superlattices,\cite{callori} is probably significantly overestimated.)

The in-plane spin-resolved conductivity in the  diffusive regime in the two polarization states is shown in Fig.\ref{fig:cond-xx-eV}. The total conductivity changes by  35\%  upon polarization switching. Also, the switching  modulates the down to up-spin conduction ratio (2.5 to 1.9)  by $\pm$15\% compared to the calculated bulk value (2.2). Such sizable modulations may be employed in in-plane field-effect devices such as that sketched in Fig.\ref{fig:cond-xx-eV}.

The electrically-stored 
polarization orientation could be read electrically  from the current modulation  in  the metal channel, i.e., effectively, as a modulated resistivity; this could be done either with the total current, or one of the spin components if polarized contacts are used. 
Another application of this configuration may be a filter or modulator of the incoming current, measuring the outgoing spin-polarized current calibrated to that of the bulk. 

Note that our calculations measure the
conduction  within the whole PTO/SRO layer system, and there is no way to single out the net contribution of each interface; the exact values of the modulation will thus depend on SRO thickness, and will change if one of the interfaces  (e.g. SRO/substrate) is ``ferroelectrically dead''. 

\section{Summary}

In conclusion, we considered a  multiferroic tunnel junction 
with asymmetric interfaces and a large-polarization FE. Very
different potential profiles result for the two polarization states, and lead to a giant TER of up to 350. The interface charge
accumulation is spin-polarized, with magnetization and magnetization changes
depending on the interfaces and   on  
 polarization orientation, with sizable ME coefficients.  The  ME coupling
affects the in-plane diffusive transport of the junction changing the majority to minority  conductivity  ratio, as well as the total conductivity. In particular, upon P inversion, the conductivity is modulated by 35\% and its spin polarization by $\pm$15\%, which is presumably exploitable in practical applications. In forthcoming work we plan to study TMR, which may be expected to be high also, as well as the effects of magnetic doping of the FE layer.

\section*{Acknowledgments} Work supported in part by MIUR-PRIN 2010 {\it Oxide}, IIT-Seed NEWDFESCM and POLYPHEMO,  IIT ``platform computation'', Fondazione Banco di Sardegna, and CINECA.


\begin{thebibliography}{References}
\bibitem{nota}
We obtain a similar value using the standard procedure of Ref.\onlinecite{key-6} for  non-polar PTO. The  monopole component could not be  completely eliminated from the potential profile in the polar case, leading to large uncertainties in the final value.
\end{thebibliography}
\end{document}